\documentclass[prl,twocolumn,showpacs,nobibnotes,floatfix,superscriptaddress]{revtex4}

\usepackage{graphicx,eucal,amsfonts}

\begin{document}

%\preprint{}
%\draft

\title{A Scalable Architecture for Coherence-Preserving Qubits}
\author{Yaakov S. Weinstein}
\thanks{To whom correspondence should be addressed}
\email{weinstein@mitre.org}
\affiliation{Quantum Information Science Group, The {\sc Mitre} Corporation, Eatontown, NJ, 07724}
\author{C. Stephen Hellberg}
\email{hellberg@dave.nrl.navy.mil}
\affiliation{Center for Computational Materials Science, Naval Research Laboratory, Washington, DC 20375 \bigskip}

\date{\today}

\begin{abstract}
We propose scalable architectures for the coherence-preserving qubits 
introduced by Bacon, Brown, and Whaley [Phys. Rev. Lett. {\bf 87}, 247902 
(2001)]. These architectures employ extra qubits providing additional 
degrees of freedom to the system. We show that these extra degrees of 
freedom can be used to counter errors in coupling strength within the 
coherence-preserving qubit and to combat interactions with environmental 
qubits. The presented architectures incorporate experimentally viable 
methods for inter-logical-qubit coupling and can implement a controlled 
phase gate via three simultaneous Heisenberg exchange operations. The 
extra qubits also provide flexibility in the arrangement of the physical 
qubits. Specifically, all physical qubits of a coherent-preserving qubit 
lattice can be placed in two spatial dimensions. Such an arrangement allows 
for universal cluster state computation.
\end{abstract}

\pacs{03.67.Lx, % Quantum computation
      03.67.Pp, % QEC and other methods for protection against decoherence
      75.10.Jm} % Quantized Spin Models
   
\maketitle

The encoding of logical qubits (LQ) into subspaces of multiple physical qubits 
is a powerful means of protecting quantum information from decoherence while 
allowing for universal quantum computation \cite{Z1,Duan}. Experimental 
examples of these decoherence free subspaces have been realized on 
nuclear magnetic resonance \cite{Evan1}, ion trap \cite{K}, and optical 
systems \cite{Alt}, have been suggested for superconducting qubits 
\cite{Zhou}, and have been used to implement encoded quantum algorithms 
\cite{Moh,Oller}. Logical qubits of this type also allow performance of 
quantum logic maximizing the use of readily available operations while 
partially or completely avoiding operations that may add complexity to the 
computing hardware or a significant amount of time to the computation. 
Specifically, this type of subspace has been suggested to perform universal 
quantum computation with only the Heisenberg exchange interaction for traditional
circuit based quantum computation \cite{Div,Levy,superco,YSW} and cluster state
computation \cite{cluster}.

The best protected LQs introduced to date are the 
coherence-preserving, or supercoherent, qubits (SQ) of Ref.~\cite{superco}. 
Supercoherent qubits, comprised of four physical qubits with equal coupling
between all pairs, minimize decoherence by establishing an energy gap 
between the logical qubit subspace and the other eigenstates of the system.
This forces all local interactions with the environment
to supply energy to the system. 
In addition, SQs allow for universal quantum computation using only the 
Heisenberg exchange coupling \cite{Div}. This increases the speed of the 
computation for quantum dot implementations
and removes the strenuous quantum hardware demands of local 
magnetic fields \cite{qdots} or $g$-factor engineering \cite{Kato} which 
would be required for single physical qubit rotations. However, a number 
of fundamental issues were left open in the original work on the SQ architecture.
Chief among them are a scalable method to couple SQs, and the physical arrangement 
of the qubits within the SQ, such that there is equal coupling  between all pairs.

In this paper, we introduce a scalable architecture for SQs with
a practical two-dimensional arrangement of the qubits.
This flexible arrangement incorporates additional degrees of freedom
which can be used to correct errors in SQ construction and unwanted 
interactions from environmental qubits. We demonstrate that these scalable 
SQs have nearly the same robustness as the SQs of Ref.~\cite{superco} against 
the dominant form of decoherence in III-V quantum dots: hyperfine coupling 
to the nuclear spins.

The inter-logical-qubit coupling in the original SQs created the most
severe obstacle to scalability. To insure the system stays in the SQ 
subspace, Ref. \cite{superco} suggests using equal couplings between 
all pairs of the eight physical qubits comprising the pair of SQs to be coupled.
This would be difficult in practice even for two SQs and the challenge would 
grow even more acute as the number of SQs is scaled up. 
 
A more practical solution was suggested in Ref. \cite{cluster}. 
If couplings between SQs are performed adiabatically \cite{YSW} the system will return to 
the logical SQ subspace after the interaction. The adiabaticity requirement 
is not difficult to achieve as adiabatic evolution is required for {\em all} 
approaches using spins in quantum dots \cite{SLM}. 

The two-SQ interaction plus equal logical $z$ rotations on each of the 
two SQs performs a conditional phase gate which, together with the single
qubit rotations of Ref. \cite{Bacon}, form a universal set of 
gates. In practice, the operations to perform the conditional phase gate 
may be done simultaneously, thus the gate requires only {\em one} time interval.

\begin{figure}
\includegraphics[width=8cm]{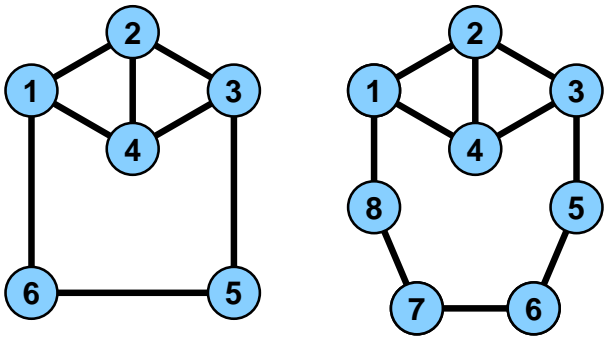} 
\caption{\label{SQ}
(Color online) Proposed layouts for supercoherent qubits (SQ) with extra 
qubits to mediate couplings. The arrangement shown on the left has
four physical qubits arranged such that the distance (and hence
the coupling) between 2 and 4 is equal to the distance between the qubits
on the edges of the rhombus.  The extra qubits 5 and 6 mediate
the coupling between 1 and 3. The right figure shows a proposed 
layout with four extra qubits mediating the coupling between 1 and 3.
Any even length mediating chain can be used to create a supercoherent qubit.}
\end{figure}

The above shows that inter-SQ interactions can be implemented
via control of the couplings between physical qubits of different SQs. 
The concern raised in \cite{superco}, that coupling between SQs will 
cause the state of the system to leave the logical SQ subspace, is 
solved by the adiabaticity of the interactions. 

Another stringent restriction in the actual construction of SQs is the 
need for equal couplings between all pairs within an SQ. Coupling 
between quantum dots is exponentially dependent on the distance 
between the electrons within the dots and can be controlled by 
electrostatic gates. A two dimensional square layout of the four physical 
qubits, such as implied in \cite{superco} and suggested in \cite{cluster}, 
is impractical because the qubits diagonally across from each other are a further 
distance apart then neighboring qubits around the perimeter of the square. Rather, 
for practical SQ realization, the four physical qubits should be arranged such 
that there is equal distance between them. An obvious possibility is to arrange 
the four qubits in a tetrahedron, thus insuring equal distance between the 
qubits. However, a tetrahedron architecture requires that the physical qubits 
be placed in three dimensions and leaves no room for flexibility in the physical 
qubit arrangement. An ideal SQ architecture would allow for the qubits to be 
arranged in only two dimensions and include additional degrees of freedom that 
would provide flexibility in the physical placement of the qubits and the couplings 
between the qubits. 

All of these issues can be solved by 
adding more physical qubits to the SQ. Additional qubits give the 
needed flexibility in the arrangement of the quantum dots while 
maintaining the energy gap between the logical subspace and the 
other states of the system. These modified SQs are immune to global 
decoherence, but are no longer immune to decoherence from single 
environmental qubits. Nevertheless, the system is exceedingly robust 
against such errors, especially when compared with previously suggested 
encodings. In addition, decoherence from a single environmental qubit can be 
combatted by modifying the strengths of couplings within the SQ. 
More importantly, the modified SQs are nearly as robust as the 
original SQs against the most important form of decoherence: 
decoherence affecting each physical qubit in an uncorrelated manner.  

Our primary proposed architecture is a six-qubit SQ with four qubits arranged 
in a rhombus, such that the distance between qubits along an edge is equal to 
the distance along the shorter of the diagonals. Two extra qubits are used to 
mediate the coupling across the longer diagonal as in the 
two dimensional arrangement shown in Fig.~\ref{SQ}a
(Any even-length chain may be used to mediate
the coupling across the longer diagonal: In Fig.~\ref{SQ}b four extra qubits
are used). We model the system with the Heisenberg Hamiltonian 
$H= J_{ij}\;{\bf S}^i\cdot{\bf S}^j$.
For ease of calculation we assume the couplings within the rhombus are equal to one. 
There are a continuum of possible values for the couplings to the
mediating qubits.  Assuming $J_{16} = J_{35}$, the coupling $J_{56}$ yielding
the degenerate singlet ground state can be shown to be
\begin{eqnarray}
J_{56} &=& \Bigl(J_{16}\sqrt{J_{16}^4+8J_{16}^3+12J_{16}^2+8J_{16}+4}
\nonumber \\
& + & J_{16}^3+4J_{16}^2-2J_{16}-4\Bigl)/(4J_{16}+4).
\end{eqnarray}
The most convenient value of the couplings can be chosen based on external 
considerations such as ease of physical layout. Importantly, we note that 
$J_{16}$ can be set greater than, less than, or equal to 1 (with a minimum 
of $\approx .85$) allowing for many possible arrangements of the physical qubits. 
Examples of SQ coupling strengths are given in Table \ref{T1}.

\begin{table}[t]
\caption{\label{T1} Coupling strengths to and between mediating 
qubits to achieve the necessary degenerate ground state in a 6-qubit 
SQ, and the size of the energy gap. All other intra-SQ couplings are 
equal to 1.} 
\begin{center}
\begin{tabular}{||c|c|c||}\hline
$J_{16} = J_{35}$ & $J_{56}$ & energy gap \\ \hline\hline
1 & $\frac{1}{8} \left(-1+\sqrt{33}\right) = 0.5931\dots $ & .1931\dots \\ \hline 
1.17672\dots & 1 & .3855\dots \\ \hline
2 & $\frac{1}{3} \left(4+\sqrt{37}\right) = 3.3609\dots $ & .8519\dots \\ \hline
\end{tabular}
\end{center}
\end{table}

The flexibility of the extra couplings allows for corrections of imperfection 
in values of other couplings. 
For example, using the coupling constants of the first 
line in Table \ref{T1}, let us assume that misplacement of qubits causes 
the value of $J_{24}$ to be 1.1 instead of 1. Due to the extra degrees of freedom 
afforded by the six-qubit SQ, this can be corrected by reducing 
the value of $J_{56}$ to approximately $.5443$. 
As another example, errors due to stray couplings between qubits 4-5 and 
4-6 can be corrected by modifying the $J_{56}$ coupling. This may be necessary 
due to the relatively close proximity of qubit 4 to qubits 5 and 6. 

\begin{figure}[t]
\includegraphics[height=5.8cm]{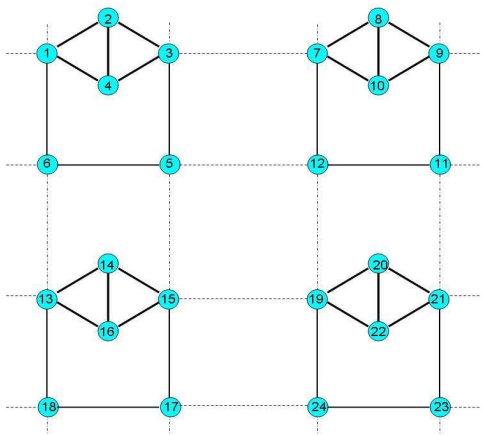}
\caption{\label{lattice}
(Color online) Arrangement of four six-dot SQs in a two-dimensional lattice. Coupling
between horizontal SQs (dashed lines) are done using the first of the coupling schemes 
described in the text and coupling between vertical SQs (chain lines) are done using 
the second scheme. All of the dots can be arranged in two spatial dimensions and universal 
cluster state quantum computation can be implemented as described in \protect\cite{cluster}.
}
\end{figure}

For the six-qubit SQ, single SQ (logical) rotations are performed 
by changing one Heisenberg coupling strength between appropriate pairs of 
qubits \cite{Bacon}. 
Depending on the choice of coupling, this performs a logical $z$ 
rotation or a rotation about the axis in the $x-z$ plane $120^{\circ}$ 
from the $z$ axis \cite{Div,YSW}. Combinations of these operations are 
sufficient to perform any $SU(2)$ rotation.

\begin{figure}[t]
\includegraphics[width=8cm]{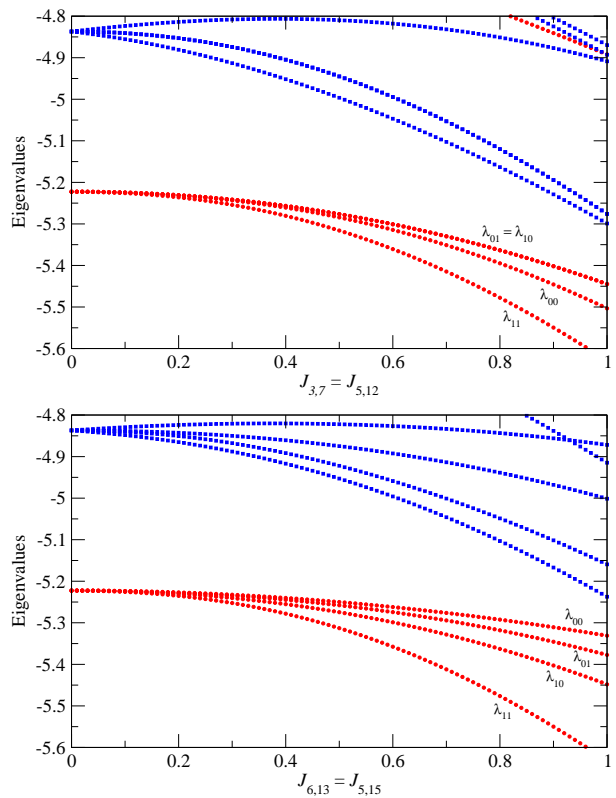}
\caption{\label{gate}
(Color online) Eigenvalues as a function of coupling strengths 
$J_{3,7} = J_{5,12}$ (top) and $J_{6,13} = J_{5,15}$ (bottom). 
These coupling schemes allow for interaction between two 
six-qubit supercoherent qubits. The $J_{3,7} = J_{5,12}$ coupling 
splits the degenerate ground state into three states spanned by the 
logical computational basis, $\lambda_{00}$, $\lambda_{11}$, the 
eigenvalues for the two-SQ logical $|00\rangle$ and $|11\rangle$ states,
and the degenerate state, $\lambda_{01}$, $\lambda_{10}$. 
The $J_{6,13} = J_{5,15}$ coupling splits the degenerate 
ground state into four states spanned by the logical computational 
basis. The above plots use coupling strengths from the second line
of Table \protect\ref{T1}. In both cases, the gap between the logical 
subspace and the rest of the system remains large even when $J_{3,7}$ or 
$J_{6,13}$ are greater than one. }
\end{figure}

While a logical qubit chain is sufficient for the implementation of universal 
circuit-based quantum computations, most algorithms can be implemented more 
efficiently on a higher dimensional lattice. Additionally a two dimensional 
lattice of logical qubits is necessary to perform universal cluster-based quantum 
computation. To this end we identify two different coupling schemes 
to perform logical operations between pairs of SQs. These are shown
in Fig.\ \ref{lattice}. Both coupling schemes result in diagonal two-logical-qubit 
gate operations. The low-lying eigenvalues of two coupled qubits are shown
in Fig.\ \ref{gate}.  For SQs coupled horizontally in Fig.\ \ref{lattice}, 
two of the eigenvalues in the computational space are equal $\lambda_{01} = \lambda_{10}$. 
For vertically coupled SQs, all four eigenvalues are different.
In both cases, the two-SQ gate operations can be combined with
single logical-qubit $z$-rotations to perform a controlled phase gate.
Since these operations commute, they may be performed simultaneously,
and the controlled phase gate can be implemented with a single pulse 
of the exchange interactions.

Other inter-six-qubit-SQ coupling methods are possible. However, the ones 
discussed above are the best we know of for satisfying the requirements 
of a diagonal inter-SQ coupling and ease of arrangement of dots.

We note that fine tuning of the couplings is necessary in order to account 
for additional interactions arising from multi-electron terms. When using a Hubbard Hamiltonian 
$ H  =  t_{ij} c_{ i \sigma}^\dagger c_{ j \sigma} + U n_{i\uparrow} n_{i\downarrow} $
for the 6-dot SQ, the degenerate ground state is obtained for hopping parameters 
$t_{ij} = \sqrt{U J_{ij}/4 }$ only in the infinite $U$ limit. For finite $U$, 
the hopping parameters need to be adjusted slightly \cite{Mizel}.

\begin{figure}[t]
\includegraphics[width=8cm]{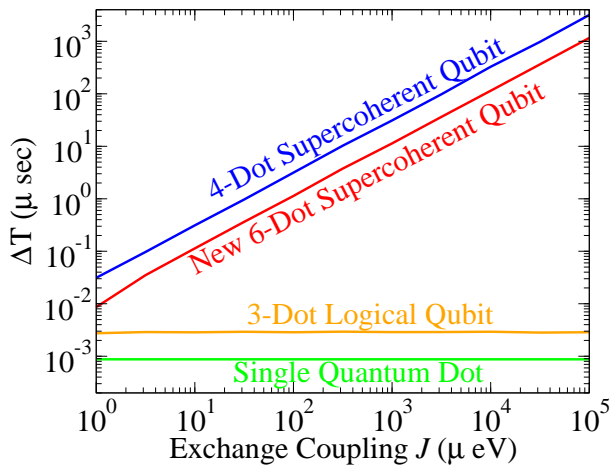} 
\caption{\label{Hb}
(Color online) Precession time of logical qubits due to the random magnetic
field of nuclei in the quantum dot as a function of 
$J$, the intra-logical-qubit coupling. When the logical qubit is just a
single qubit or consists of three physical qubits, as in \protect\cite{YSW}, 
the precession time is independent of $J$. For the four-qubit 
and six-qubit SQs the precession time increases 
linearly with $J$. 
The figure shows the dramatic improvement in decoherence time
gained by properly encoding qubits. The magnetic field strength used is 
.06 $\mu$eV. This and other quantities are based on 
the experimental work in \protect\cite{Petta}.
}
\end{figure}

The dominant source of decoherence in III-V quantum dots is the local random 
magnetic field generated by the nuclei in each dot \cite{Sasha}. The
effect on a SQ from this type of decoherence is a splitting of the 
degenerate ground state ({\em i.e.}\ the logical qubit subspace), 
by $\Delta E$, leading to a precession, or decoherence, time given by 
$\Delta T = \hbar/ (2\pi\Delta E)$. The precession times for the 
SQ containing 4 and 6 physical qubits are shown in Fig. \ref{Hb} as a 
function of $J$, the coupling between two physical qubits within a given logical
qubit (for the 6-qubit case the ratio given in the first line of Table \ref{T1} 
was used). Also shown are the precession times of a single qubit and the 3-dot logical
qubits of Ref.\ \cite{YSW}. We note that current experimental data suggests a ratio 
between the strength of the effective magnetic field, 
$H_b$, and $J$ of $10^{-6} \alt H_b/J \alt .01$ \cite{Petta}. Immediately 
noticeable is the orders of magnitude increase in decoherence
time for the 4- and 6-qubit SQs. In addition, upon increasing $J$, the decoherence 
time for both of these increases linearly and the six-qubit case loses little 
in robustness against this type of decoherence when compared with the four-qubit SQ. 

In conclusion, we have introduced flexible architectures for supercoherent
qubits with the goal of reducing the severe constraints required for 
equal inter-qubit couplings. These constraints include placing qubits
at exact distances from each other which forces the placing of qubits in 
three dimensions. The schemes introduced here increase flexibility while 
keeping the energy gap necessary to protect the SQ from sources of decoherence. 
The additional degrees of freedom can be used to correct mismatches in intra-SQ 
couplings and to reduce coupling from environmental qubits. The SQs can be connected 
in both one and two-dimensional arrangements, and their natural implementation of diagonal 
logical operations makes them particularly suitable for cluster-state quantum computation.
Most importantly, the supercoherent qubits show a dramatic increase in robustness 
against decoherence due to nuclei, the primary source of decoherence in III-V quantum dots.

The authors would like to thank Al.L.Efros for stimulating discussions. Y.S.W.
acknowledges the support of MITRE Technology Program Grant 07MSR205. C.S. Hellberg
acknowledges support from the DARPA QuIST program.

\end{document}